\documentclass[prd,twocolumn,nofootinbib,utf8]{revtex4}

\usepackage{amssymb}
\usepackage{amsfonts}
\usepackage{setspace} 
\usepackage{graphicx} 
\usepackage{amsthm}
\usepackage{gensymb}
\usepackage{mathrsfs} 
\usepackage{mathtools}
\usepackage{amsmath}
\usepackage{hyperref} 
\hypersetup{
    colorlinks=true,
    linkcolor=blue,
    filecolor=blue,      
    urlcolor=blue,
    citecolor=blue,
    }

\begin{document} 

\title{A pedagogical approach to the black hole information issue} 

\author{Thiago T. Bergamaschi}
\email{tbergamaschi@ifsc.usp.br}
\affiliation{Sao Carlos Institute of Physics, University of Sao Paulo, IFSC – USP,
13566-590, Sao Carlos, SP, Brazil}

\begin{abstract}

We provide a pedagogical introduction to the concepts underlying black hole information loss, intended for readers familiar with special relativity and quantum mechanics. We emphasize that there is no paradox of information loss and that proposals that deviate from well-established theories in arbitrary regimes are inherently contradictory.
  
   \vspace{.1cm}
   \noindent
   \textbf{Keywords}: General relativity. Quantum field theory in curved spacetime. Semiclassical gravity. Hawking radiation. Black hole information issue.
     
\end{abstract}

\maketitle

%%%%%%%%%%%%%%%%%%%%%%%%%%%%%%%%%%%%%%%%%%%%%%%%%%%%%%%%%%%%%%%%%%%%%%%%%%%%%%%%%%%%%%%%%%%%%%%%%%%%%%%%%%%%%%%%%%%%%
\section{Introduction}
%%%%%%%%%%%%%%%%%%%%%%%%%%%%%%%%%%%%%%%%%%%%%%%%%%%%%%%%%%%%%%%%%%%%%%%%%%%%%%%%%%%%%%%%%%%%%%%%%%%%%%%%%%%%%%%%%%%%%
The concept of a region from which nothing can escape due to gravity, known as a \textit{black hole}, has been a notable front for developments in theoretical physics over the last few decades. One of the most remarkable outcomes of these developments is the state of affairs commonly referred to as the \textit{black hole information paradox}. This unfortunate nomenclature stems mainly from an \textit{erroneous} interpretation of the implications of black holes for information preservation, which has led to many \textit{physically unreliable} proposals. To fully grasp the nature of this issue and understand the associated misconceptions, it is beneficial to begin with a review of the current descriptions of the fundamental interactions.

\begin{figure}[h]
\centering
\includegraphics[scale=1.3]{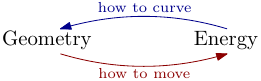} 
\caption{Qualitative description of the gravitational interaction proposed by GR.}
\label{1}
\end{figure}

\textit{General relativity} (GR) \cite{Wald1984} is a theory of the gravitational interaction whose characterization is made by the \textit{curvature} of \textit{spacetime} due to the presence of an \textit{energy distribution}. A qualitative notion of this description is given in fig. \ref{1}, inspired by the quote \cite[p.~5]{Misner1973}: ``\textit{Spacetime tells matter how to move; matter tells spacetime how to curve}''. This \textit{classical}\footnote{By classical, it is meant that the characterization of the gravitational interaction, given by the spacetime geometry, assumes definite values in any length scale, not only those much larger than atomic ones (i.e., distances of order $10^{-9}\;\text{m}$).} theory currently paints the most accurate picture of gravitational phenomena, and thus, is the one we use to describe black holes. On the other hand, the present depiction of the \textit{electroweak} and \textit{strong} interactions is given by \textit{quantum field theory} (QFT) \cite{Parker2009}, which can be understood as an extension of \textit{quantum mechanics} (QM) and \textit{special relativity} (SR) that treats \textit{quantized fields} as protagonists of these interactions. Being highly successful at describing observational data and experiments in both their pertinent regimes of application, these theories can be combined in the mathematically rigorous but physically incomplete formalism of \textit{quantum field theory in curved spacetime} (QFTCS).

\begin{figure}[h]
\centering
\includegraphics[scale=1.2]{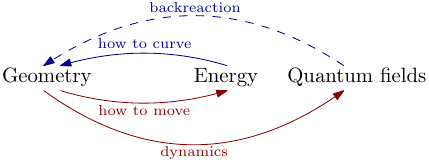} 
\caption{Qualitative description of the interactions in QFTCS. The dashed line indicates that, within this framework, there is no general description of backreaction effects.}
\label{2}
\end{figure}

The argument behind the physical incompleteness of QFTCS follows from GR, which quantifies how a classical energy distribution, described by the \textit{energy-momentum tensor}, affects the spacetime geometry. In QFTCS, the quantum field dynamics is governed by the curved spacetime, but one does not know how to generally describe the effects of the quantum field energy content on the spacetime geometry, known as \textit{backreaction effects} (see fig. \ref{2}). Nonetheless, one can rely on physical arguments to \textit{qualitatively} describe possible backreaction for specific phenomena, which leads to the more complete framework of \textit{semiclassical gravity} (SG)\footnote{Although we distinguish between QFTCS and SG due to the consideration of backreaction effects, some references use these terms as synonyms.}. Consequently, following the experimental success of GR and QFT, SG provides the best available simultaneous description of the fundamental interactions. In other words, SG provides the best formalism available to study quantum effects in and \textit{on} gravitation. 

A notable quantum effect is the \textit{Hawking effect}, which states that stationary (i.e., time independent) black holes effectively emit\footnote{Hawking radiation does not ``get out'' from a black hole.} an \textit{approximate thermal spectrum of particles}, known as \textit{Hawking radiation}. Now, if one considers the implication of a negative energy flux entering the black hole due to local energy-momentum conservation, then the main backreaction effect due to Hawking radiation will be to reduce the black hole mass and other physical parameters, resulting in an effective \textit{evaporation} process. Hence, after a finite time, a black hole will reach the Planck scale, in the sense that its radius will be of the order of the \textit{Planck length}, 
\begin{equation}
    \ell_p=\left(\frac{G\hbar}{c^3}\right)^{1/2}\approx 10^{-35}\;\text{m},
\end{equation}
where $G$ is Newton's constant, $\hbar$ is Planck's constant, and $c$ is the speed of light. However, the spacetime curvature at such a regime will be extreme, and dimensional arguments suggest that the smooth spacetime structure postulated by GR should cease to be an adequate description. In this sense, one expects that a full quantum treatment of \textit{gravitational degrees of freedom} and spacetime structure should be necessary for an accurate picture (i.e., a theory of \textit{quantum gravity}) at the Planck scale, as semiclassical predictions will no longer be sensible. 

To understand how black holes lead to information loss in SG, consider the following line of reasoning. Suppose one starts with a stationary energy distribution, which is described by its mass, angular momentum, electric charge, baryon and lepton number, and any other physical property. From the perspective of QM, complete knowledge of the quantum state of this distribution would correspond to a \textit{pure} state. An example of this would be a highly diluted cold-matter distribution (e.g., a dust cloud) such that one has all the possible information about its configuration. Of course, such a distribution would not be exactly stationary due to gravitation and other interactions, but, in distant regions, it would be a good approximation to describe it by a stationary spacetime. Now, as this distribution evolves, it could form, say, a star. If such a star is sufficiently massive, at the later stages of its life, gravitational effects will not be sustained, and a black hole will form as a consequence of \textit{gravitational collapse}. Even though the black hole formation process is expected to be highly dependent on the energy distribution details and how it collapses, asymmetries of the distribution are expected to be radiated away in the form of \textit{gravitational waves}. Following a conjecture that black holes must eventually reach a stationary configuration, the initial energy distribution will then give rise to a black hole described only by three parameters: its mass, angular momentum, and electric charge. However, the Hawking effect predicts that such black holes should behave as gray bodies, effectively emitting an approximate thermal spectrum of radiation. This radiation carries away the black hole mass, angular momentum, and electric charge, and after a finite time, the black hole might completely \textit{evaporate}, leaving only radiation. In light of SG, the resulting quantum state would necessarily correspond to a \textit{mixed} one, which arises in a context in which one does not have all the possible information about the state. Therefore, the complete evaporation of a black hole would take a pure state to a mixed one, corresponding to the fact that most information about the details of the energy distribution that gave rise to the black hole is lost.

Evidently, for this conclusion to hold, one would have to trust SG predictions at arbitrary scales. Indeed, even though SG predicts an evaporation process for regimes far from the Planck scale, when a black hole reaches a radius of order $\ell_p$, only a theory of quantum gravity would be appropriate to describe its fate accurately. Consequently, \textit{it is an interesting open question how a theory of quantum gravity would address black hole evaporation}. Nonetheless, the conclusion of information loss in the framework of SG \textit{contradicts no known phenomena and established theories}; in fact, it is a \textit{physical prediction}. In contrast, it has been extensively and \textit{incorrectly} argued in the literature that the conclusion of information loss is \textit{paradoxical} in light of \textit{unitary evolutions} in QM. As a result, many \textit{inconsistent} proposals have come to light to solve a \textit{nonexistent} paradox. 

The purpose of this paper is twofold. First, we provide a pedagogical presentation of the concepts that lead to the conclusion that information is lost due to black holes in SG. Second, we show why there is no paradox of information loss and discuss the flaws in proposals that seek to invalidate SG predictions at arbitrary regimes. The paper is organized as follows. In sec.~\ref{classical}, we discuss black holes in the framework of GR, introducing the necessary ideas for their definition and analysis. In sec.~\ref{semiclassical}, we present the Hawking effect and the implication of quantum entanglement for black holes in SG. In sec.~\ref{loss}, we show how SG predicts that black holes lead to information loss and discuss misconceptions surrounding this conclusion. We conclude in sec.~\ref{con} with the final remarks. We assume familiarity with SR and QM. The arguments presented are discussed in much more detail in \cite{Bergamaschi2024}, which includes an extensive pedagogical review of the technical analysis required to support the ideas of this paper. We work with SI units, but we set $c=1$ for all spacetime diagrams\footnote{This is a mere visual simplification to allow null curves to be aligned with lines of slope $\pm1$.}. Unless stated otherwise, we will consider four-dimensional spacetimes.

%%%%%%%%%%%%%%%%%%%%%%%%%%%%%%%%%%%%%%%%%%%%%%%%%%%%%%%%%%%%%%%%%%%%%%%%%%%%%%%%%%%%%%%%%%%%%%%%%%%%%%%%%%%%%%%%%%%%%
\section{Classical aspects of black holes}\label{classical}
%%%%%%%%%%%%%%%%%%%%%%%%%%%%%%%%%%%%%%%%%%%%%%%%%%%%%%%%%%%%%%%%%%%%%%%%%%%%%%%%%%%%%%%%%%%%%%%%%%%%%%%%%%%%%%%%%%%%%

In the purely classical framework of GR, a black hole is defined as a region of spacetime from which nothing can escape. In this section, we detail the classical concepts necessary to make this notion precise and explain the importance of stationary black holes. Our starting point is to discuss pertinent properties of Minkowski spacetime.

\subsection{Minkowski spacetime}\label{min}

In SR, the background for all phenomena is the \textit{Minkowski spacetime}, which is a collection of events endowed with 
a four-dimensional measure (i.e., the \textit{invariant interval}) to every pair of such events given by the \textit{Minkowski metric}. The concept of \textit{causality} follows from the possible events that can influence one another, say, by a massive particle or an electromagnetic wave. The analysis of which events can be causally connected then defines the \textit{causal structure} of Minkowski spacetime, which is given by the set of events that can be interpreted as the \textit{future} and \textit{past} of events.

\begin{figure}[h]
\centering
\includegraphics[scale=1.5]{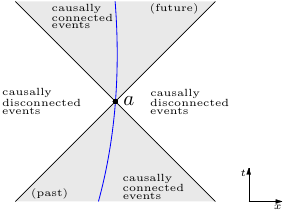} 
\caption{Future and past of an event, $a$, in two-dimensional Minkowski spacetime. The illustration contains a timelike curve (blue) and null curves (black).}
\label{3}
\end{figure}

For instance, consider an event in Minkowski spacetime, $a$. Associated to $a$ are two sets of events: those that can be connected to it by a \textit{timelike} or \textit{null} curve\footnote{These curves must obey pertinent differentiability conditions and be oriented in a single direction of time (see \cite[sec.~2.2]{Bergamaschi2024} for more details).}, and those that cannot. The latter is the set of events that are \textit{causally disconnected} from $a$, while the former is the set of events that are \textit{causally connected} to $a$. In this manner, the notion of future and past for causally connected events arises from the fact that either a particle or wave can be emitted to influence $a$ or to have emanated from $a$. These ideas are illustrated in fig. \ref{3}. Note that any event inside the gray region is causally connected to $a$, as well as those along the null curves that pass through $a$. Events not in the gray regions or on its boundary are causally disconnected from $a$, meaning that only faster-than-light travel would allow one to reach them from $a$. In particular, the distinction between the future and past sets is given by a predetermined continuous definition of what constitutes going ``forwards'' or ``backwards'' in time. All curves considered in the following should be understood as ``going'' from the past to the future.

\begin{figure}[h]
\centering
\includegraphics[scale=1.2]{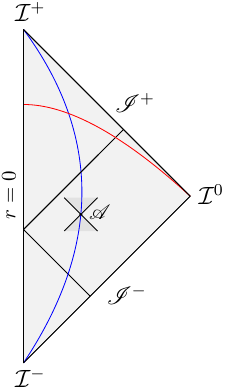} 
\caption{Conformal diagram of Minkowski spacetime with two angular coordinates suppressed. The illustration contains null curves (black), a timelike curve (blue), and a three-dimensional spacelike section (red).}
\label{4}
\end{figure}

A helpful way to visualize the entire causal structure of Minkowski spacetime is by a \textit{conformal diagram} (also known as a \textit{Carter-Penrose diagram}). Such a diagram allows one to depict the entire spacetime in a finite illustration, while keeping null curves aligned with lines of slope $\pm1$. The conformal diagram of Minkowski spacetime is illustrated in fig. \ref{4}. In the coordinate system $\{(t,r,\theta,\phi)\}$, where $t$ is the time parameter of static inertial observers in the usual spherical coordinates $\{(r,\theta,\phi)\}$, the interpretation of the conformal diagram is as follows\footnote{In the following, all conformal diagrams will be presented with two angular coordinates suppressed.}. The vertical black line is the center of the coordinate system, $r=0$, for different values of $t$. All null curves begin at the three-dimensional null hyperplane $\mathscr{I}^-$ and end at the three-dimensional null hyperplane $\mathscr{I}^+$. Similarly, all timelike curves begin at the point $\mathcal{I}^-$ and end at the point $\mathcal{I}^+$. Additionally, the red line is a three-dimensional spacelike section (e.g., a $t=\text{constant}$ hyperplane) that begins at $r=0$ and ends at $\mathcal{I}^0$. A diagram analogous to that of fig. \ref{3} is also represented in fig. \ref{4}, where the null curves emanating from the two-sphere, $\mathscr{A}$, should be understood to extend until they reach $\mathscr{I}^-$ and $\mathscr{I}^+$. Consequently, the future and past of any other two-sphere can be readily verified by tracing null curves through it, i.e., delimiting its \textit{light cone}. 

\begin{figure}[h]
\centering
\includegraphics[scale=1.2]{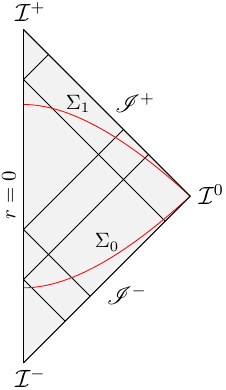} 
\caption{Conformal diagram of Minkowski spacetime illustrating two Cauchy surfaces (red), $\Sigma_0$ and $\Sigma_1$.}
\label{5}
\end{figure}

One of the most notable features of the conformal diagram of Minkowski spacetime are the two null hyperplanes $\mathscr{I}^+$ and $\mathscr{I}^-$. These are known as \textit{future null infinity} and \textit{past null infinity}, respectively. They can be interpreted as ``infinitely distant regions'' of Minkowski spacetime, such that only light rays can reach, to the future and to the past (i.e., asymptotic future and past). This is because all null curves begin at $\mathscr{I}^-$ and end at $\mathscr{I}^+$. To visualize this, consider a light ray that was emitted far away from the origin of the coordinate system (i.e., $r\gg0$) at early times (i.e., $t\ll0$). If the light ray was emitted so that its radial coordinate decreases over time (i.e., $dr/dt=-c$), then it will eventually reach the origin of the coordinate system, $r=0$. After that, its radial coordinate will increase over time (i.e., $dr/dt=c$), and it will eventually reach the regions far away from the origin at late times (i.e., $t\gg0$). The regions $\mathscr{I}^-$ and $\mathscr{I}^+$ appear in this line of reasoning when one takes the limit of $r\to\infty$ (i.e., far away region) and $t\to\pm\infty$ (i.e., early and late times). In particular, the process of the light ray reaching $r=0$ and thus being described by a different null curve is illustrated by the ``reflection'' of the null curve in fig. \ref{4} (see also fig. \ref{5} for more examples).

To conclude our analysis, we mention an important property of Minkowski spacetime: it possesses a \textit{Cauchy surface}\footnote{A Cauchy surface is a closed hypersurface which is intersected by each inextendible timelike and null curve exactly once (see \cite[sec.~2.2]{Bergamaschi2024} for more details).}. To understand this in more detail, consider any three-dimensional spacelike section that begins at $r=0$ and extends to $\mathcal{I}^0$, such as $\Sigma_0$ and $\Sigma_1$ in fig. \ref{5}. It is possible to see that all timelike and null curves intersect $\Sigma_0$ and $\Sigma_1$ exactly once. For instance, null curves emerge from $\mathscr{I}^-$ and intersect $\Sigma_0$ and $\Sigma_1$ exactly once, either on their way to reach $r=0$ or on their way to reach $\mathscr{I}^+$ after ``reflecting'' at $r=0$. Thus, both $\Sigma_0$ and $\Sigma_1$ are Cauchy surfaces. Spacetimes possessing a Cauchy surface are said to be \textit{globally hyperbolic}, which can be interpreted as those that obey pertinent conditions on \textit{causality} and \textit{determinism}, allowing one to formulate a \textit{well-posed problem} for dynamics.

Globally hyperbolic spacetimes can be described as a \textit{foliation} of Cauchy surfaces, i.e., as a collection of Cauchy surfaces. An example of this is a foliation of Minkowski spacetime by $t=\text{constant}$ hyperplanes. Although a foliation is not unique, in a given foliation, each Cauchy surface can be interpreted as an ``instant of time'' of spacetime. Essentially, information on a Cauchy surface can be used to determine the entire development of spacetime, i.e., physical conditions on any other Cauchy surface. Indeed, because each null curve must intercept a Cauchy surface exactly once, information can be interpreted to be carried over (i.e., conserved) from one Cauchy surface to the others. This can be exemplified by considering both $\Sigma_0$ and $\Sigma_1$, as all null curves that intersect $\Sigma_0$ will eventually intersect $\Sigma_1$. Conversely, complete predictability from physical conditions at some ``instant of time'' will not hold for spacetimes that are not globally hyperbolic. Consequently, it is challenging to envision how information could be preserved when analyzing non-globally hyperbolic spacetimes.

%%%%%%%%%%%%%%%%%%%%%%%%%%%%%%%%%%%%%%%%%%%%%%%%%%%%%%%%%%%%%%%%%%%%%%%%%%%%%%%%%%%%%%%%%%%%%%%%%%%%%%%%%%%%%%%%%%%%%
\subsection{Definition}
%%%%%%%%%%%%%%%%%%%%%%%%%%%%%%%%%%%%%%%%%%%%%%%%%%%%%%%%%%%%%%%%%%%%%%%%%%%%%%%%%%%%%%%%%%%%%%%%%%%%%%%%%%%%%%%%%%%%%
In GR, the flat geometry of Minkowski spacetime is merely a particular solution for the spacetime geometry when no energy distribution is present. Following the description of fig. \ref{1}, spacetime is now postulated to \textit{curve} due to the presence of an energy distribution. Nevertheless, the notion of a causal structure is still pertinent to a curved spacetime, in the sense that one can rely on the same arguments as those presented in Minkowski spacetime to define the future and past of events. Our focus now is to use these ideas to give a rigorous definition of a black hole.

The intuitive notion of a black hole is a region from which nothing can escape. To make this notion precise, one needs to describe what it means ``to escape from'' and to where it could ``escape''. Notably, the idea that there is a restriction on the behavior of timelike and null curves is precisely captured by the concept of causally disconnected events. Namely, if there is no null or timelike curve connecting two events, then they are causally disconnected. Consequently, ``to escape from'' can be understood as the possibility of events being causally connected. Additionally, the idea of where null curves could contemplate ``escaping to'' is given by future null infinity, $\mathscr{I}^+$. As discussed in sec. \ref{min}, in Minkowski spacetime, all null curves end at $\mathscr{I}^+$, which means that, naturally, one expects that null curves would eventually reach $\mathscr{I}^+$, ``escaping from'' the rest of spacetime. In this sense, the definition of a black hole can be made precise in spacetimes possessing a region similar to $\mathscr{I}^+$. These are known as \textit{asymptotically flat spacetimes}, as the ``infinitely distant regions'' have a Minkowski-like character. In other words, the asymptotic regions to the past and the future in asymptotically flat spacetimes behave similarly to $\mathscr{I}^-$ and $\mathscr{I}^+$.

\begin{figure}[h]
    \centering
    \includegraphics[scale=1.2]{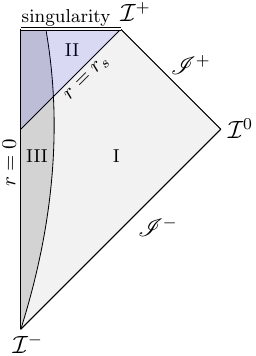}
    \caption{Conformal diagram of Schwarzschild spacetime illustrating the exterior vacuum region (I), the black hole region (II) (highlighted in blue), and the collapsing energy distribution (III). Note that the black hole region contains part of region III, and also an interior vacuum region.}
    \label{6}
\end{figure}

In this manner, \textit{the black hole region of an asymptotically flat spacetime is defined as the set of events that are causally disconnected from $\mathscr{I}^+$}. Physically, this can be interpreted as the fact that no information emanating from regions inside the black hole can reach $\mathscr{I}^+$. In other words, nothing can escape from the black hole due to gravity, not even light. An example of this is given in fig. \ref{6}, which illustrates the conformal diagram of a \textit{spherically symmetric} and electric \textit{uncharged} energy distribution that collapsed to form a black hole, known as a \textit{Schwarzschild black hole}. The spacetime describing such a black hole is known as \textit{Schwarzschild spacetime}, and the interpretation of its diagram is the following. Region II, highlighted in blue, is the \textit{black hole region}. Region I is the exterior \textit{vacuum} region, i.e., no classical energy distribution is present. Region III is the interior region of the spherically symmetric energy distribution that collapses to form a black hole. Region II is delimited by the radial coordinate defined by
\begin{equation}\label{eq2}
    r=\left(\frac{A}{4\pi}\right)^{1/2},
\end{equation}
where $A$ is the black hole area. During the collapse, $A$ goes from zero to a maximum, such that the maximum value of the radial coordinate in eq. \ref{eq2} is given by the \textit{Schwarzschild radius},
\begin{equation}\label{eq3}
    r_s=\frac{2GM}{c^2}\approx 10^{3} \left(\frac{M}{M_\odot}\right)\; \text{m},
\end{equation} where $M$ is the energy distribution total mass and $M_\odot$ is the Sun's mass. The boundary of the black hole region is known as the \textit{event horizon}. The thick white line at the top of the diagram represents the \textit{singularity} inside the Schwarzschild black hole, a ``pathological'' region of spacetime. The rest of the diagram boundary has the same interpretation as that of the conformal diagram of Minkowski spacetime.

\begin{figure}[h]
    \includegraphics[scale=1.2]{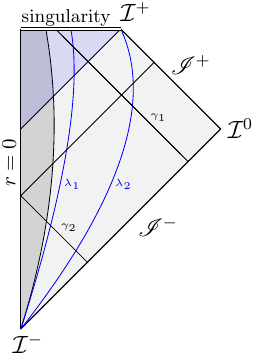}
    \caption{Conformal diagram of Schwarzschild spacetime illustrating pairs of timelike curves (blue) and null curves (black).}
    \label{7}
\end{figure}

To visualize why region II is the black hole region, consider some examples of null and timelike curves, as shown in fig. \ref{7}. The timelike curve $\lambda_2$ is an example of the worldline of a massive particle that stays outside region II, and that will end at $\mathcal{I}^+$. In contrast, the timelike curve $\lambda_1$ is an example of the worldline of a massive particle that enters region II, and its worldline ends not at $\mathcal{I}^+$, but rather, at the singularity. Similarly, the null curve $\gamma_2$ reaches $\mathscr{I}^+$, while $\gamma_1$ enters region II and eventually reaches the singularity. Indeed, any event in region II will not only be unable to communicate with events outside of it, but any massive particle, electromagnetic wave, or signal emanating from it will unavoidably reach the singularity. Note that this also holds for the energy distribution in region III. More precisely, after it collapses (i.e., reaches $r_s$), all the distribution will eventually reach the singularity.

\begin{figure}[h]
    \centering
    \includegraphics[scale=1.2]{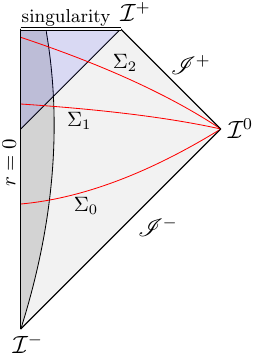}
    \caption{Conformal diagram of Schwarzschild spacetime illustrating three Cauchy surfaces (red), $\Sigma_0$, $\Sigma_1$, and $\Sigma_2$.}
    \label{fig30}
\end{figure}

\begin{figure}[h]
    \centering
    \includegraphics[scale=1]{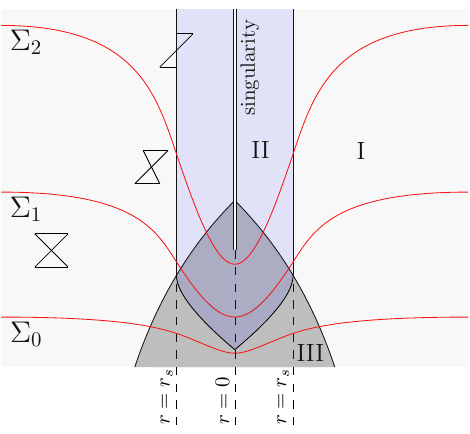}
    \caption{Sketch of the \textit{Finkelstein diagram} corresponding to the conformal diagram shown in fig. \ref{fig30}. In this diagram, only null curves whose radial coordinate decreases over time are aligned with lines of slope $\pm1$, as shown by the ``tilting'' of the light cones as they approach $r=r_s$.}
    \label{fig31}
\end{figure}

In this manner, defining the black hole region as the set of events causally disconnected from $\mathscr{I}^+$ captures the idea of a region from which nothing can escape. However, the fact that $\mathscr{I}^+$ represents an asymptotic region means that one would require complete knowledge of spacetime to analyze which events are causally connected to it. Due to this, it is convenient to restrict the definition of black holes to spacetimes that are globally hyperbolic at least outside the black hole and on the event horizon. With this, one can study the \textit{dynamical aspects} of black holes in a covariant manner using Cauchy surfaces. Indeed, from figs. \ref{fig30} and \ref{fig31}, it is easy to see that Schwarzschild spacetime is globally hyperbolic by using the same criterion as for Minkowski spacetime (see fig. \ref{5}). On the other hand, without this condition of global hyperbolicity, it is difficult to perceive a way to discuss a \textit{physically significant} notion of time evolution of black holes. 

%%%%%%%%%%%%%%%%%%%%%%%%%%%%%%%%%%%%%%%%%%%%%%%%%%%%%%%%%%%%%%%%%%%%%%%%%%%%%%%%%%%%%%%%%%%%%%%%%%%%%%%%%%%%%%%%%%%%%
\subsection{Stationary black holes}
%%%%%%%%%%%%%%%%%%%%%%%%%%%%%%%%%%%%%%%%%%%%%%%%%%%%%%%%%%%%%%%%%%%%%%%%%%%%%%%%%%%%%%%%%%%%%%%%%%%%%%%%%%%%%%%%%%%%%
Stationary black holes are those that do not change over time. Their analysis is pertinent mainly for two reasons. First, one expects that all \textit{physical} black holes (i.e., those resulting from \textit{gravitational collapse})\footnote{In the following, we will only consider physical black holes.} should be eventually described, to a good order of approximation, by a time independent configuration. Second, theorems in GR state that any stationary black hole can be described by only three parameters. 

The first reason stems from expectations about the product of gravitational collapse which results in a black hole. In particular, although details of the gravitational collapse can greatly affect the spacetime geometry, at sufficiently ``late times'' after its formation (as ``measured'' by Cauchy surfaces), a black hole is expected to reach a stationary configuration due to interaction with other sources of energy and the emission of \textit{gravitational waves} (i.e., undulations of curvature propagating in spacetime). This \textit{conjecture} is corroborated by the behavior of other systems, e.g., electromagnetic ones, in which a time dependent distribution of electric charges radiates away the higher-order multipole moments and eventually ``settles down'' to a stationary distribution. 

The second reason comes from the results known as the \textit{black hole uniqueness theorems}. These theorems show that any stationary black hole must be described uniquely by its \textit{mass}, $M$, \textit{angular momentum}, $L$, and \textit{electric charge}, $q$. The most general stationary black hole solution is known as a \textit{Kerr-Newman black hole}, described by the parameters $(M,L,q)$. Evidently, a Schwarzschild black hole is a \textit{nonrotating} (i.e., $L=0$) and electric uncharged (i.e., $q=0$) Kerr-Newman black hole. Coupled with the conjecture that black holes should eventually reach a stationary configuration, the black hole uniqueness theorems lead to the conclusion that any black hole will eventually have only three degrees of freedom. 
%%%%%%%%%%%%%%%%%%%%%%%%%%%%%%%%%%%%%%%%%%%%%%%%%%%%%%%%%%%%%%%%%%%%%%%%%%%%%%%%%%%%%%%%%%%%%%%%%%%%%%%%%%%%%%%%%%%%%
\section{Semiclassical aspects of black holes}\label{semiclassical}
%%%%%%%%%%%%%%%%%%%%%%%%%%%%%%%%%%%%%%%%%%%%%%%%%%%%%%%%%%%%%%%%%%%%%%%%%%%%%%%%%%%%%%%%%%%%%%%%%%%%%%%%%%%%%%%%%%%%%
In this section, we will discuss the two main quantum effects that are pertinent to the conclusion of information loss in black holes. The first concerns the effective particle creation by black holes, known as the Hawking effect. The second is about an implication of entanglement between regions of spacetime in the framework of SG. Our starting point is to explain the concept of particle content in QFT.

\subsection{Hawking effect}
In QFT, a \textit{quantized field} is the protagonist, being described by a collection of \textit{quantum harmonic oscillators} in spacetime. In this framework, what one would classically refer to as \textit{particles} are now viewed as \textit{quanta} or \textit{excitations} of the oscillators describing the quantum field. As one would proceed in QM, the evaluation of a quantum state \textit{particle content} is done by the computation of the expectation value of the \textit{number operator}. Consequently, dynamical processes that generate excitations of the quantum field can be understood as resulting in \textit{particle creation}. This is the fundamental concept behind the Hawking effect.

\begin{figure}[h]
\centering
\includegraphics[scale=0.8]{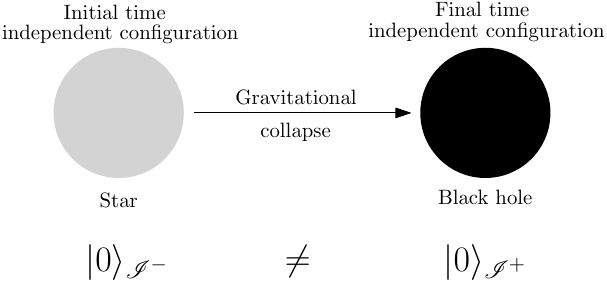} 
\caption{Creation of particles due to the process of gravitational collapse and black hole formation \cite{Fabbri2005}.}
\label{8}
\end{figure}

Essentially, an observer ``near'' $\mathscr{I}^-$ would identify an initial \textit{no particle state} (i.e., one with no quanta or excitations) as the \textit{vacuum state}, $|0\rangle_{\mathscr{I}^-}$. An example of such an observer is one at the early stages of its life, with a worldline given by $\lambda_1$ or $\lambda_2$ in fig. \ref{7}. Now, let $|0\rangle_{\mathscr{I}^+}$ denote the vacuum state for an observer ``near'' $\mathscr{I}^+$. An example of such an observer is one that remains outside the black hole at the late stages of its life, with a worldline also given by $\lambda_2$ in fig. \ref{7}. Notably, due to gravitational collapse and black hole formation, one finds that $|0\rangle_{\mathscr{I}^-}$ does not simply evolve to $|0\rangle_{\mathscr{I}^+}$, as illustrated in fig. \ref{8}. This means that particles are created as witnessed by observers in the asymptotic future. 

The content of the particle spectrum measured at $\mathscr{I}^+$ can be evaluated to be \textit{predominantly} proportional to that of a \textit{black body}, whose temperature is given by the \textit{Hawking temperature},
\begin{equation}\label{temp}
    T=\frac{\hbar \kappa}{2\pi ck_B},
\end{equation}
where $k_B$ is Boltzmann's constant and $\kappa$ is the black hole \textit{surface gravity}. For a Schwarzschild black hole, the surface gravity reads
\begin{equation}\label{kappa}
    \kappa=\frac{c^4}{4GM}\approx 10^{13}\left(\frac{M}{M_\odot}\right)^{-1}\; \text{m/s$^2$}.
\end{equation}
This result can be interpreted as stating that a Kerr-Newman black hole will, effectively, behave predominantly as a gray body\footnote{The more detailed form of the spectrum contains \textit{gray body factors} due to transmission probabilities.}, whose temperature is proportional to its surface gravity\footnote{Although the local acceleration necessary to keep a massive particle still at the event horizon is divergent, the surface gravity is a finite quantity. This is because it is defined as the acceleration as measured from $\mathscr{I}^+$.}. This effective particle creation by black holes is known as the \textit{Hawking effect}, while the approximate thermal radiation predicted by it is referred to as \textit{Hawking radiation}. As expected from the uniqueness theorems, a detailed evaluation of the spectrum for a Kerr-Newman black hole reveals that the emission of particles depends only on the parameters $(M,L,q)$. Also, the emission of particles with angular momentum and electric charge with the same sign as those of the black hole has a higher probability than those with opposite sign. 

%%%%%%%%%%%%%%%%%%%%%%%%%%%%%%%%%%%%%%%%%%%%%%%%%%%%%%%%%%%%%%%%%%%%%%%%%%%%%%%%%%%%%%%%%%%%%%%%%%%%%%%%%%%%%%%%%%%%%
\subsection{Entanglement}
%%%%%%%%%%%%%%%%%%%%%%%%%%%%%%%%%%%%%%%%%%%%%%%%%%%%%%%%%%%%%%%%%%%%%%%%%%%%%%%%%%%%%%%%%%%%%%%%%%%%%%%%%%%%%%%%%%%%%
Entanglement is a feature that arises in \textit{composite quantum systems}, limiting one's ability to obtain complete information about an \textit{individual system}. Namely, suppose one is dealing with the two-particle state 
\begin{equation}
    |\psi\rangle=\frac{1}{\sqrt{2}}|\uparrow_1\rangle\otimes|\downarrow_2\rangle+\frac{1}{\sqrt{2}}|\downarrow_1\rangle\otimes|\uparrow_2\rangle.
\end{equation}
Evidently, this is a \textit{pure} composite state, because it is represented by a single ket in the total \textit{Hilbert space}. Additionally, it is an \textit{entangled} state, as it cannot be written as a simple product state in the individual Hilbert spaces. Indeed, by ignoring the degrees of freedom of system $1$ (i.e., by ``tracing-out'' its degrees of freedom in the state $|\psi\rangle$) one finds that system $2$ is in a \textit{mixed} state described by the \textit{density operator}
\begin{equation}
    \hat{\rho}=\frac{1}{2}|\uparrow_2\rangle\langle\uparrow_2|+\frac{1}{2}|\downarrow_2\rangle\langle\downarrow_2|.
\end{equation}
Notably, mixed states cannot be written as a ket in the Hilbert space. This means that, by restricting one's attention only to system $2$, there is now a \textit{probability distribution} of states, and one cannot know for certain what state describes the individual system. In this sense, entanglement affects how much information one can acquire about the state of a single subsystem.

In the context of SG, entanglement is an essential feature in light of what one would expect to be a \textit{physically acceptable class of states}. In particular, since one is ultimately interested in describing \textit{backreaction} effects (see fig. \ref{2}), it is of significance to understand properties that a state should possess to make sense of a notion of \textit{energy content}. To understand this in more detail, consider the following line of reasoning.

Recall that the ground state for a quantum harmonic oscillator has energy $\hbar\omega/2$. If one considers the infinite collection of quantum harmonic oscillators that describe a quantum field, then an associated vacuum state has a \textit{divergent energy content}. Indeed, the nontrivial structure of the vacuum state in QFT leads to several effects that have been experimentally observed. However, no problems arise in the theoretical framework due to procedures that address divergent quantities (i.e., \textit{regularization} and \textit{renormalization}). For example, in Minkowski spacetime, one performs a \textit{vacuum energy subtraction} procedure using the natural vacuum state defined in it, the \textit{Minkowski vacuum}, $|0\rangle_M$. This procedure allows one to give a finite notion of the energy content of a state, at least for states belonging to a certain class.

Yet, such a procedure is not valid for a general curved spacetime because there is no analog of the Minkowski vacuum. Nevertheless, one can select a class of states that satisfy the condition of having a divergent character similar to that of the Minkowski vacuum. States obeying this property are known as \textit{Hadamard states}, and it can be argued that for a state to be physically acceptable, it must be a Hadamard state. Indeed, it is possible to define a procedure that results in a finite energy content for Hadamard states \cite{Wald1994, Parker2009}. 

\begin{figure}[h]
\centering
\includegraphics[scale=1.2]{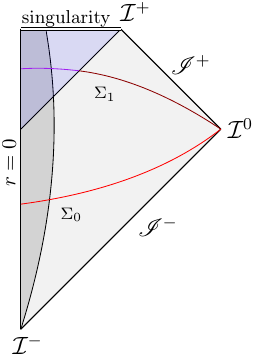} 
\caption{Conformal diagram of Schwarzschild spacetime illustrating two Cauchy hypersurfaces, $\Sigma_0$ (red) and $\Sigma_1$ (divided in two portions, dark red (outside the black hole) and purple (inside the black hole)).}
\label{9}
\end{figure}

One of the main features of Hadamard states is that they imply \textit{entanglement over spacelike separated regions}. In fig. \ref{9}, we illustrate how this is of significance for black holes using two Cauchy surfaces, $\Sigma_0$ and $\Sigma_1$. Assuming that the initial state is a pure Hadamard state, then the field state at $\Sigma_0$ is also pure. The dynamical evolution of physical conditions from $\Sigma_0$ to any other Cauchy hypersurface will be unitary, and thus, will maintain its purity. For instance, for the ``instant of time'' $\Sigma_1$, the \textit{total} field state is also pure. But, because $\Sigma_1$ intersects the black hole region, observers \textit{outside} the black hole will have access only to a portion of $\Sigma_1$ (the dark red portion). Consequently, due to the entanglement implied by the Hadamard state, the field state \textit{inside} the black hole ``at a time'' $\Sigma_1$ (the purple portion) will be entangled with the one outside. Analogously to the spin case, the state outside the black hole will be mixed after one traces out degrees of freedom inside the event horizon. Evidently, they can always blame the black hole for concealing information about the quantum field, in the sense that information about the state inside the black hole could only be accessed once one enters it. Nevertheless, information about the state inside the black hole will never be able to be shared with observers outside the black hole. In other words, even though the spacetime is globally hyperbolic, observers outside the black hole will experience a mixed state simply because they do not have access to the black hole interior.

%%%%%%%%%%%%%%%%%%%%%%%%%%%%%%%%%%%%%%%%%%%%%%%%%%%%%%%%%%%%%%%%%%%%%%%%%%%%%%%%%%%%%%%%%%%%%%%%%%%%%%%%%%%%%%%%%%%%%
\section{Time evolution of semiclassical black holes}\label{loss}
%%%%%%%%%%%%%%%%%%%%%%%%%%%%%%%%%%%%%%%%%%%%%%%%%%%%%%%%%%%%%%%%%%%%%%%%%%%%%%%%%%%%%%%%%%%%%%%%%%%%%%%%%%%%%%%%%%%%%
In this section, we show how the semiclassical properties of black holes discussed so far imply that information is lost. We will show why this conclusion is in agreement with semiclassical arguments, and does not earn the adjective ``paradoxical''. Additionally, we will present misconceptions about proposals that seek to change the conclusion regarding information loss.

%%%%%%%%%%%%%%%%%%%%%%%%%%%%%%%%%%%%%%%%%%%%%%%%%%%%%%%%%%%%%%%%%%%%%%%%%%%%%%%%%%%%%%%%%%%%%%%%%%%%%%%%%%%%%%%%%%%%%
\subsection{Information loss}
%%%%%%%%%%%%%%%%%%%%%%%%%%%%%%%%%%%%%%%%%%%%%%%%%%%%%%%%%%%%%%%%%%%%%%%%%%%%%%%%%%%%%%%%%%%%%%%%%%%%%%%%%%%%%%%%%%%%%

Although calculating backreaction effects is not a trivial task for four-dimensional spacetimes, one expects that the main consequence of the Hawking radiation is to reduce the black hole parameters at a given time, i.e., its mass, angular momentum, and electric charge. This results in an effective \textit{evaporation} process. This \textit{qualitative} description of a possible backreaction effect is corroborated by physical arguments. More precisely, the energy reduction is merely a consequence of the fact that the positive energy flux to $\mathscr{I}^+$ implies that there must exist a negative energy flux going into the black hole, while the reduction of the other parameters is because the spectrum of emitted particles tends (i.e., is more likely) to carry away angular momentum and electric charge. Because of the spectrum characteristics, a Kerr-Newman black hole will quickly (in comparison with time scales of interest) become a Schwarzschild black hole. Therefore, for simplicity, our analysis will be focused on a Schwarzschild black hole.

As per eqs. \ref{temp} and \ref{kappa}, the Hawking temperature of a Schwarzschild black hole is inversely proportional to its mass, which means that the typical energy of created particles increases as a Schwarzschild black hole evaporates. In particular, this relation between particle energy spectrum and black hole mass ensures that created particles will have a small influence on the spacetime metric for most of the evaporation process, so that neglecting additional backreaction effects will be a good approximation when $r_s\gg\ell_p$ (i.e., far from the Planck scale). In this sense, details of backreaction effects will have little influence over the event horizon and the exterior region. Consequently, the spacetime geometry can be described by a sequence of \textit{quasi-static processes} in which the black hole radius decreases slowly, with the process of energy loss being approximated by \textit{Stefan's law}. This argument leads one to the conclusion that a black hole should completely evaporate, i.e., radiate away its mass, in a finite amount of time (as ``measured'' by Cauchy hypersurfaces). Evidently, at some point in time $r_s\sim\ell_p$, hence, complete evaporation is only a valid prediction if one considers that no deviations from SG occur at that scale. For the moment, let us consider that to be the case and analyze the consequences of a complete evaporation process. 

\begin{figure}[h]
\centering
\includegraphics[scale=1.2]{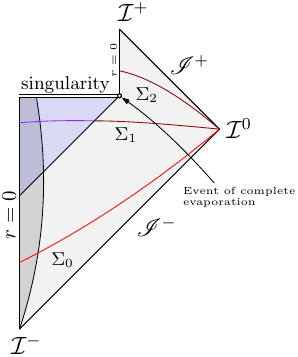} 
\caption{Conformal diagram of a Schwarzschild black hole that forms from a spherically symmetric energy distribution and evaporates completely.}
\label{10}
\end{figure}

The conformal diagram of a Schwarzschild black hole that results from a spherically symmetric collapse and completely evaporates is illustrated in fig. \ref{10}. Notably, \textit{the spacetime fails to be globally hyperbolic}. This can be verified by the fact that some null curves do not intersect $\Sigma_2$. More precisely, although after complete evaporation events at $r=0$ are again causally connected to $\mathscr{I}^+$, there are null curves that never intersect $\Sigma_2$ because they reach the singularity\footnote{Note that the event of complete evaporation corresponds to a singularity that is causally connected to $\mathscr{I}^+$. Thus, null curves emanating from it do not intersect any section such as $\Sigma_0$ or $\Sigma_1$.}. Hence, physical conditions at $\Sigma_2$ will not suffice to determine those at $\Sigma_0$ or $\Sigma_1$. \textit{Fundamentally, information loss can be attributed to the fact that some null curves reach the singularity.} Indeed, information loss is also evident from the fact that the field state at a spacelike section such as $\Sigma_2$ (``after'' complete evaporation) is not sufficient to determine the complete field state at either $\Sigma_0$ or $\Sigma_1$. In particular, considering the same color scheme of fig. \ref{9}, at $\Sigma_0$ one has an initial pure Hadamard state. At $\Sigma_1$, entanglement due to the Hadamard state means that the field state outside the black hole will be a mixed one. Consequently, \textit{after complete evaporation}, the field state at any spacelike section such as $\Sigma_2$ will also be a mixed one.

The generality of this analysis follows from the arguments presented in secs. \ref{classical} and \ref{semiclassical}. Namely, suppose one starts with an isolated energy distribution described by a pure Hadamard state. If this distribution is sufficiently massive, gravitational effects will not be sustained, and a black hole will form. Following the emission of gravitational waves, the black hole exterior will eventually be well approximated by a Kerr-Newman metric. The Hawking effect, together with the qualitative description of backreaction, indicates that the Kerr-Newman black hole will (quickly) result in a Schwarzschild black hole. One is then led to the same line of reasoning presented above, where information about the field state after complete evaporation does not suffice to determine the initial state.

\begin{figure}[h]
\centering
\includegraphics[scale=0.85]{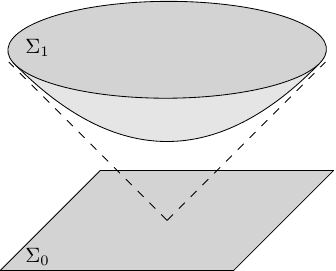} 
\caption{Spacetime diagram of Minkowski spacetime illustrating a Cauchy hypersurface, $\Sigma_0$, and an asymptotically null hyperboloid, $\Sigma_1$ \cite{Wald1994}.}
\label{11}\end{figure}

Now, the expectation of information conservation is only justified in a globally hyperbolic spacetime, where information is carried from one Cauchy surface to another. In fact, QFT predicts that information on a Cauchy surface will not, in general, be completely deposited in a future region that fails to be a Cauchy surface. This means that even in a globally hyperbolic spacetime, one can still find examples of ``information loss'' (in the sense that a pure state evolves to a mixed one). For instance, in Minkowski spacetime, one can consider the field evolution from a Cauchy surface, $\Sigma_0$, to an asymptotically null hyperboloid, $\Sigma_1$ (see fig. \ref{11}) \cite{Wald1994}. Because there are null curves that do not intersect $\Sigma_1$, a pure state at $\Sigma_0$ would necessarily evolve to a mixed one at $\Sigma_1$. As such, dynamical evolution from $\Sigma_0$ to $\Sigma_1$ would not be unitary as it does not constitute a \textit{closed} system (i.e., an isolated system). Evidently, this is not at odds with the postulates of QM, which require unitary evolutions only for closed systems.

Similarly, information loss due to the formation and \textit{complete} evaporation of black holes can be interpreted as a consequence of the singularity resulting in an \textit{open} system (i.e., a non-isolated system). In this sense, the prediction of information loss in SG is a \textit{physical} one, and hence, \textit{should not be interpreted as paradoxical}. 
%%%%%%%%%%%%%%%%%%%%%%%%%%%%%%%%%%%%%%%%%%%%%%%%%%%%%%%%%%%%%%%%%%%%%%%%%%%%%%%%%%%%%%%%%%%%%%%%%%%%%%%%%%%%%%%%%%%%%
\subsection{Misconceptions about alternatives to information loss}

It has been extensively argued in the literature that a pure-to-mixed-state evolution violates physical principles and contradicts conjectures about black holes (e.g., that black holes are ordinary thermodynamic systems \cite{Almheiri2021}), and therefore black hole evaporation must be a unitary process. The \textit{incorrect} line of reasoning that leads to these interpretations views information loss as paradoxical, and refers to the state of affairs surrounding the product of black hole evaporation as the \textit{black hole information paradox}. 

Notably, if one proposes a different description of black hole evaporation that leads to information conservation, \textit{it is imperative to preserve semiclassical physics far from the Planck scale}. More precisely, a proposal that seeks to change semiclassical predictions at \textit{arbitrary} regimes is \textit{inherently contradictory}. Examples of such contradictory proposals are those that support the view that black holes never form (e.g., \cite{Mathur2005}) and those arguing that Hawking radiation should strongly deviate from thermality (e.g., \cite{Almheiri2021}). Note that formation of black holes and the derivation of the predominant thermal character of the Hawking radiation are robust results that do not rely on specific regimes. For instance, consider the curvature scales at the event horizon of a Schwarzschild black hole, which are proportional to $r_s^{-2}$. From eq. \ref{eq2}, to avoid the formation of \textit{any} Schwarzschild black hole, one would have to find deviations from SG predictions in arbitrary curvature scales. The exact same argument applies to deviations from thermality by Hawking radiation. That is, a strong deviation from thermality at arbitrary regimes could only be accounted for by strong deviations from the semiclassical picture far from the Planck scale. However, \textit{QFT and GR are well-tested at low curvature regimes} (e.g., solar system-like curvature settings), and, because no deviations are found, relying on arguments that would require such descriptions to fail at arbitrary regimes is contradictory. On the other hand, nothing can be said about proposals that change the product of the evaporation process only at the Planck scale, neither in favor nor against them. 

\section{Conclusions}\label{con}

In this paper, we reviewed the classical and semiclassical descriptions of black holes and presented the conjectures and hypotheses that lead to the conclusion of information loss. We show that the \textit{physical} non-unitary dynamical evolution that follows from black hole complete evaporation can be interpreted as a consequence of the singularity ``opening'' the system. Consequently, \textit{the conclusion of information loss in SG is not paradoxical}. Additionally, we discussed the contradictory nature of proposals that suggest deviations from SG at arbitrary scales. 

While acknowledging that a theory of quantum gravity is required to make precise predictions about black hole evaporation, particularly regarding backreaction effects and Planck-scale physics, we emphasize that the hypotheses, conjectures, and approximations of SG are well justified. Nonetheless, it remains a compelling open issue how black hole evaporation will be addressed within an adequate theory of quantum gravity, and what novel physics it could provide. 

\section*{Acknowledgments}

The author thanks Esmerindo de Sousa Bernardes for his support during the early stages of this work, Daniel A. T. Vanzella for extensive support and comments on a draft, and George E. A. Matsas for many conversations on black hole physics and support during the later stages of this work. The author also thanks two anonymous referees for their detailed and helpful comments. This study was financed by the Coordena\c c\~ao de Aperfei\c coamento de Pessoal de N\'ivel Superior – Brasil (CAPES) – Finance Code 001, and by the Sao Paulo Research Foundation (FAPESP), Brazil. Process Number 2024/19057-7.

\end{document}